\documentclass{article}
\usepackage{graphicx}
\usepackage[a4paper, margin=1in]{geometry}
\usepackage{amsmath}
\usepackage{tcolorbox}
\usepackage{url}
\usepackage{hyperref}
\usepackage[backend=biber]{biblatex}

\title{High Performance Matrix Multiplication}
\author{Ethan Davis}
\date{March 2025}

\begin{document}

\maketitle

\section{Introduction}

Matrix multiplication is the foundation from much of the success from high performance technologies like deep learning, scientific simulations, and video graphics. High level programming languages like Python and R rely on highly optimized low level libraries for performing core linear algebra operations like matrix multiplication from Basic Linear Algebra Subprograms (BLAS). This paper compares the performance of five different matrix multiplication algorithms using CuBLAS, CUDA, BLAS, OpenMP, and C++ Threads. We find statistical significance with a p-value below 5e-12 to support the hypothesis that for square $N \times N$ matrices where $N$ is at least $10,000$ then the in order performance as measured in floating point operations per second (FLOPS) for these matrix multiplication algorithms is CuBLAS, CUDA, BLAS, OpenMP, and C++ Threads.

\section{Methods}

In this section we describe how we measure the performance of different matrix multiplication algorithms. In Subsection \ref{subsection:Experiment} we outline our approach for data collection and analysis and in Subsection \ref{subsection:Metrics} we define the metrics of our data. Subsections \ref{subsection:Hardware} and \ref{subsection:Software} describe our hardware specifications including runtime environment and high level pseudocode for our hand rolled algorithms respectively. Lastly, Subsections \ref{subsection:Host} and \ref{subsection:Device} describe how our hand rolled algorithms are implemented for CPU and GPU processing. Our C++ and CUDA implementations along with our Python scripts used for statistical analysis are available on our GitHub \cite{davis-2025} with documentation to repeat our experiments.

\subsection{Experiment} \label{subsection:Experiment}

\sloppy
We compare five matrix multiplication algorithms using CuBLAS, CUDA, BLAS, OpenMP, and C++ Threads. For each algorithm we calculate matrix multiplication $30$ times with ten square $N \times N$ matrix sizes where $N$ equals 1e3, 2e3, 3e3, 4e3, 5e3, 6e3, 7e3, 8e3, 9e3, and 1e4. That is, matrix multiplication is calculated $30$ times for each pair of algorithm and matrix size. For each matrix multiplication we use randomly generated matrices with double precision floating point elements sampled from the inclusive range $[2.0, 5.0]$ with uniform distribution. With $30$ trials for each pair of algorithm and matrix size we calculate the sample mean FLOPS using the bootstrap method and include $95\%$ confidence intervals between the $2.5$ and $97.5$ percentiles.

Our algorithms take advantage of cache locality. For all matrix multiplication calculations we use square $K \times K$ tile sizes where $K = 32$. Later we analyze in depth why this tile size is optimal for all algorithms and our hardware. Since we use GPU devices in our experiment there is time spent copying memory between the host and device. However, we exclude this time spent copying from our runtime and calculated FLOPS measurements. That is, matrix multiplication is measured by time spent for arithmetic calculation alone.

Our goal is to determine whether our algorithms have the same performance as measured in FLOPS. Our hypothesis is that some algorithms have better performance than others. We use the null hypothesis that FLOPS for all algorithms are the same. Our alternative hypothesis is that FLOPS for some algorithms are different.

\subsection{Metrics} \label{subsection:Metrics}

We use FLOPS to measure performance for all of our matrix multiplication algorithms. Given square $N \times N$ matrices there is a dot product for each row and column of the left and right matrix respectively. For each dot product there are $N$ scalar multiplications and $N-1$ scalar additions. Therefore, the exact number of floating point operations for matrix multiplication of square matrices is $2N^3 - N^2$. We use this exact number of floating point operations and divide by the number of seconds needed to complete matrix multiplication to calculate FLOPS.

\begin{equation*}
N^2 \times (N + (N - 1)) =
N^2 \times (2N - 1) =
2N^3 - N^2.
\end{equation*}

\subsection{Hardware} \label{subsection:Hardware}

Our CuBLAS and CUDA algorithms rely on host CPU and device NVIDIA GPU resources. Our BLAS, OpenMP, and C++ Threads algorithms only use CPU resources. Tables \ref{tab:HostSpecs} and \ref{tab:DeviceSpecs} give the specifications of our host and device used for all matrix multiplication calculations. Both our Intel Xeon Gold 5118 and NVIDIA Tesla V100 processors belong to a class of server level processors as opposed to being designed for personal workloads \cite{intel-2017, nvidia-2017}. For our experiments, all matrix multiplication calculations were completed directly on hardware and without the use of hypervisors or virtual machines and without resource sharing for other processes.

\begin{table}
\centering
\begin{tabular}{|c|c|}
\hline
Property & Value \\
\hline
Architecture & Intel Xeon Gold 5118 \\
Cores & 48 \\
Base Clock Speed & 1000 MHz \\
Max Clock Speed & 3200 MHz \\
L1 Cache & 32 KB \\
L2 Cache & 1024 KB \\
L3 Cache & 16 MB \\
Memory & 200 GB \\
Operating System & Debian GNU/Linux 10 \\
Compiler & g++ 8.3.0 \\
\hline
\end{tabular}
\caption{Host specifications \cite{intel-2017}.}
\label{tab:HostSpecs}
\end{table}

\begin{table}
\centering
\begin{tabular}{|c|c|}
\hline
Property & Value \\
\hline
Architecture & NVIDIA Tesla V100 \\
Threads / Warp & 32 \\
Max Thread Block Size & 1024 \\
SM Count & 80 \\
FP64 Cores / SM & 32 \\
Base SM Clock Speed & 1245 MHz \\
Max SM Clock Speed & 1380 MHz \\
Shared Memory / SM & 64 KB \\
Register File Size / SM & 256 KB \\
L2 Cache & 6144 KB \\
Global Memory & 16 GB \\
Compiler & NVCC 11.2 \\
\hline
\end{tabular}
\caption{Device specifications \cite{nvidia-2017}.}
\label{tab:DeviceSpecs}
\end{table}

\subsection{Software} \label{subsection:Software}

\begin{figure}[h!]
    \centering
    \includegraphics[width=0.7\textwidth]{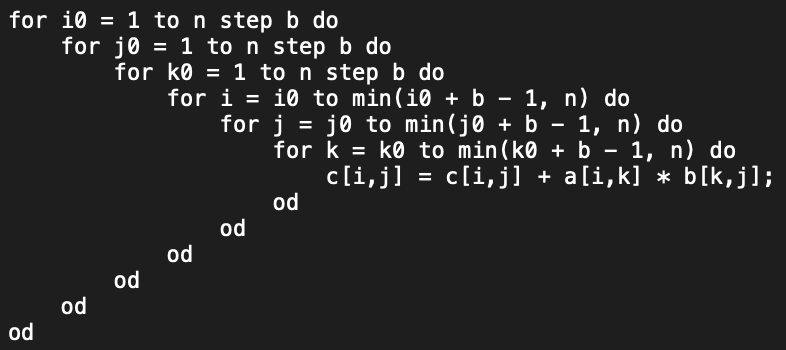}
    \caption{Tiled matrix multiplication pseudocode \cite{dongarra-1997}.}
    \label{fig:Pseudocode}
\end{figure}

CuBLAS and BLAS are third party libraries that we benchmark against our hand rolled CUDA, OpenMP, and C++ Threads algorithms. For our hand rolled algorithms we use tiled matrix multiplication pseudocode outlined in Figure \ref{fig:Pseudocode}. Matrix multiplication has time complexity $O(N^3)$ since for each row and column of the left and right matrices respectively there is a dot product. This pseudocode lets us take advantage of cache locality by breaking down large matrices into small tiles. We complete matrix multiplication calculation between these tiles that fit in cache resources and therefore reduce the number of memory accesses.

For square $K \times K$ tile size $K$, each element of the left and right tiles is accessed $K$ times. By choosing $K$ such that tiles fit in cache resources, we reduce the number of memory accesses by a factor of $K$ since each element is only accessed from memory once. Using tiled matrix multiplication as an optimization increases our floating point operation to memory access ratio. For a row $i$ and column $j$ of the left and right matrices respectively, dot product results only need to be written to memory once after all tiled dot products are calculated.

The pseudocode steps through rows $i0$ and columns $j0$ of tiles with sequence $k0$. For each tile, a tiled matrix multiplication is performed with rows $i$ and columns $j$ using elements in sequence $k$. The iteration of rows $i0$ and columns $j0$ can be completed in parallel. That is because calculations are written to unique memory addresses. However iterations of $k0$ must be completed in sequence to avoid race conditions because these calculations are written to the same addresses.

\begin{figure}[h!]
    \centering
    \includegraphics[width=0.8\textwidth]{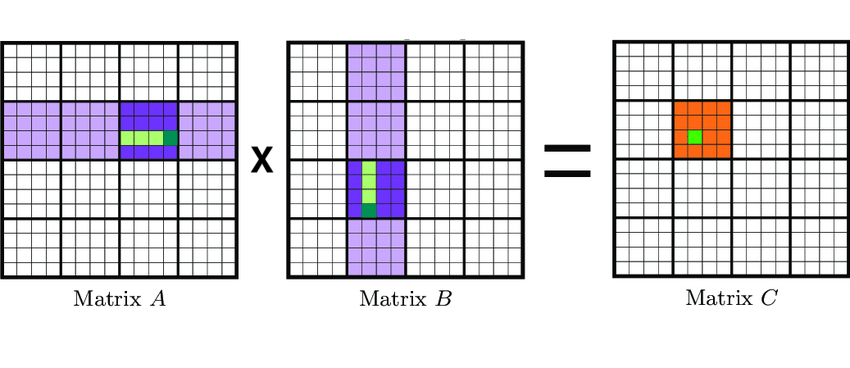}
    \caption{Tiled matrix multiplication. The light purple is iteration $k0$ and the dark purple is its current step. The light green is iteration $k$ and the dark green is its current element.}
    \label{fig:Gemm}
\end{figure}

\subsection{Host} \label{subsection:Host}

For our OpenMP and C++ Threads algorithms only the host CPU is used for calculation. Our desire is to keep tiles from the left, right, and product matrices $\mathbf{A}$, $\mathbf{B}$, and $\mathbf{C}$ respectively in cache resources. That is, we want to keep $3 \times K^2 \times 8$ bytes in our cache where $3$ number of $K \times K$ tiles of double precision floating point elements ($8$ bytes each) are kept.

Each CPU core has its own L1 cache, L2 cache shared with one other core, and L3 cache shared with all cores. Table \ref{tab:HostSpecs} shows that our host has $48$ CPU cores and each core has $32$ KB of L1 cache. By choosing tile size $K = 32$ we require $3 \times 32^2 \times 8 = 24,576$ KB of cache resources or less than $32$ KB. Therefore, our tiled matrix multiplication completes all calculation with L1 cache accesses. This optimizes our use of cache locality and floating point operation to memory access ratio.

Our OpenMP algorithm simply implements the tiled matrix multiplication pseudocode in C++. Then, using OpenMP, we collapse the two outermost loops of $i0$ and $j0$. Doing so assigns sequential iterations of inner loops $k0$, $i$, $j$, and $k$ to unique CPU cores for parallel tiled matrix multiplication calculation. Additional optimizations used are the use of local variables to hold the terminal condition of inner loops $i$, $j$, and $k$ to avoid redundant calculations. We also use a local variable to hold the running sum of the dot product from iteration $k$ to avoid redundant writes to the product matrix $\mathbf{C}$ in memory. That is, we only write to memory after the tiled dot product is complete. Since iterations $k0$ are performed sequentially we avoid race conditions when we write tiled calculations to memory.

For our C++ Threads algorithm we also parallelize the two outermost iterations $i0$ and $j0$. Using a task queue, we load all tiles as tasks to the queue. Then, we launch a thread pool where threads synchronize unloading tasks from the queue and then calculate iterations $k0$, $i$, $j$, and $k$ in parallel. This algorithm also uses the same local variables as above to avoid redundant calculations and writes to memory.

\subsection{Device} \label{subsection:Device}

Our CUDA algorithm uses the same square $K \times K$ tile size where $K = 32$. That is, $32 \times 32 = 1024$ is our CUDA block size. Recalling Table \ref{tab:DeviceSpecs}, we see that the maximum thread block size is $1024$ and so this design choice maximizes thread occupancy in our blocks. This allows our kernel to take advantage of latency hiding \cite{hwu-2022}. When a resident warp needs to wait for some long latency operation such as global memory access, another resident warp that is not waiting for results is selected for execution. GPU streaming multiprocessors (SMs) achieve zero overhead scheduling by holding all execution states for assigned warps in hardware registers so there is no need to save and restore states when switching from one warp to another.

Table \ref{tab:DeviceSpecs} shows that our GPU has $32$ threads per warp and $32$ double precision floating point (FP64) cores per SM. Therefore, our blocks with $1024$ threads are broken down into the maximum $32$ warps. Our oversubscription of threads to SMs is essential for latency tolerance. It increases the chances of finding another warp to execute when a currently executing warp encounters a long latency operation. The ratio of the number of warps assigned to an SM to the maximum number it supports is the definition of occupancy \cite{hwu-2022} which our chosen block size maximizes.

Our kernel uses shared memory for tiled matrix multiplication. This increases locality of tiles while they're being used for calculation and increases our floating point operations to global memory access ratio. Table \ref{tab:DeviceSpecs} shows that our device has $64$ KB of shared memory per SM. We store tiles from the left and right matrices in shared memory and that requires $2 \times 32^2 \times 8 = 16,384$ KB or below $64$ KB limit. Furthermore, our device has $256$ KB register file size per SM. Therefore, our kernel design avoids register spill to local memory. Lastly, since our largest matrix size is $N = 10,000$ we require at most $3 \times 10,000^2 \times 8 = 2.4$ GB of global memory which is below our limit of $16$ GB for our device.

Our CUDA algorithm performs iterations of $i0$ and $j0$ from our pseudocode in parallel. That is, each block is responsible for the calculation of one tile. For each sequential iteration $k0$, tiles from the left and right matrices are loaded into shared memory. Our kernel design is fine grained rather than coarse grained where each thread is responsible for one cell of a matrix tile as opposed to multiple cells. Our kernel can afford for each thread to process calculations for one cell in parallel as opposed to multiple cells in sequence because our resource usage is below its bounded limits. Before any calculation starts, we must synchronize all threads. That is because our algorithm relies on shared memory being loaded so that all elements of tiles can be accessed for calculation. Additionally, we must synchronize all threads again after calculation completes before loading subsequent tiles since this ensures that all threads have finished calculation.

In order to handle square matrices of any size, we must introduce control divergence into our CUDA algorithm. For matrix sizes that are not evenly divisible by our tile size, there are blocks that have unused threads. Control divergence is undesirable because it reduces occupancy. Conditional logic is used to load shared memory with matrix elements and for tile cells that do not have matrix elements we load zero. This is a safe approach because it leaves dot products unaffected. We use control divergence again for boundary checks when we write dot product calculations to global memory. Until then calculations are kept in the registers assigned to a thread.

\begin{figure}[h!]
    \centering
    \includegraphics[width=0.5\textwidth]{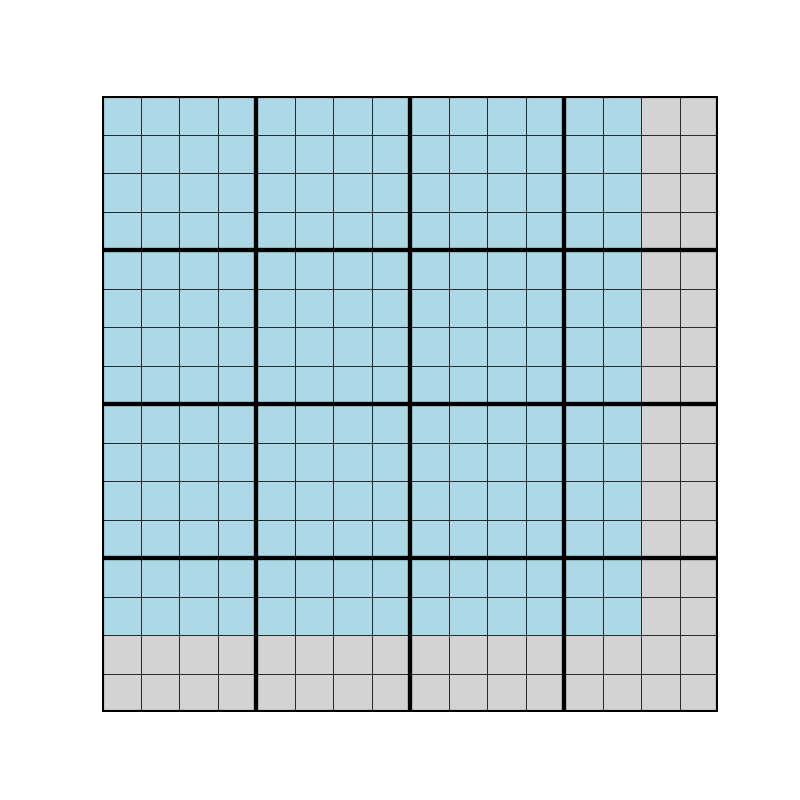}
    \caption{Tiled matrix where the matrix size is not evenly divisible by the tile size.}
    \label{fig:Tiles}
\end{figure}

Our CUDA kernel also makes use of memory coalescing. Our matrices are stored in global memory with row major order. When rows from the matrices are loaded into shared memory from global memory, these consecutive locations are accessed and delivered in a single combined (coalesced) request with DRAM bursts \cite{hwu-2022}. This occurs because consecutive capacitors in DRAM that hold individual bits as electrical charges can share in sequence with sensors when a transistor is used to drive a line to outlet gates.

\section{Results}

\begin{figure}[h!]
    \centering
    \includegraphics[width=0.7\textwidth]{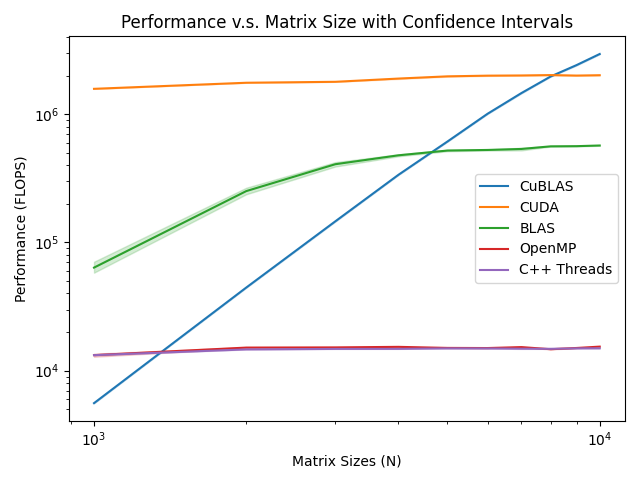}
    \caption{Performance of all algorithms plotted in logarithmic scale.}
    \label{fig:LogLog}
\end{figure}

\begin{figure}[h!]
    \centering
    \includegraphics[width=0.7\textwidth]{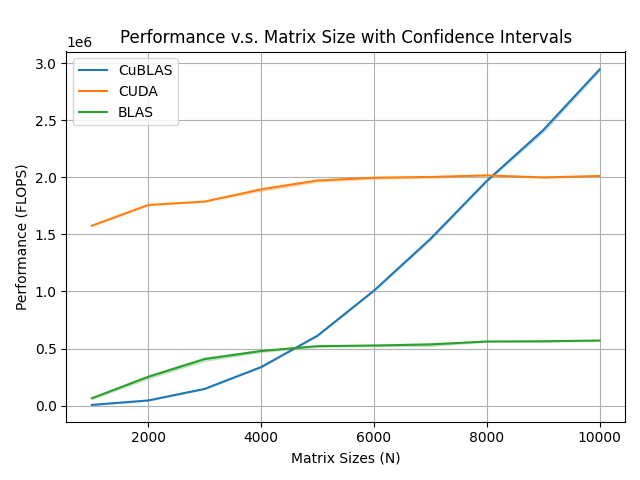}
    \caption{CuBLAS, CUDA, and BLAS performance.}
    \label{fig:GPU}
\end{figure}

\begin{figure}[h!]
    \centering
    \includegraphics[width=0.7\textwidth]{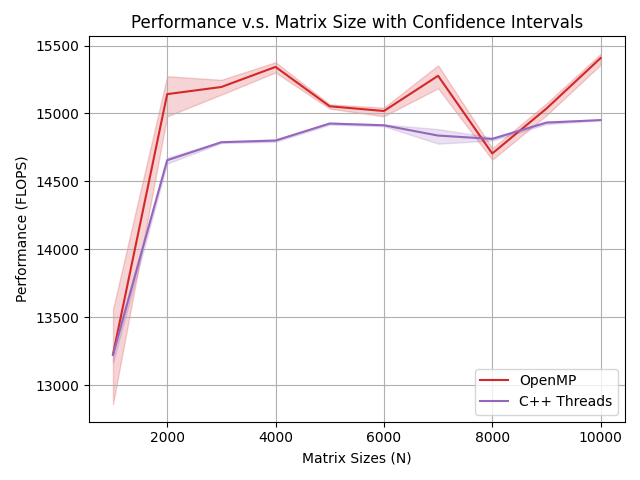}
    \caption{OpenMP and C++ Threads performance.}
    \label{fig:CPU}
\end{figure}

For large matrix sizes our GPU bases CuBLAS and CUDA algorithms outperformed our CPU based OpenMP and C++ Threads algorithms by multiple orders of magnitude. Our CPU based BLAS also outperformed OpenMP and C++ Threads by an order of magnitude. Figure \ref{fig:LogLog} show this comparison with all five algorithms plotted with logarithmic scale. For more detailed plots we separate our CuBLAS, CUDA, and BLAS algorithms into one plot and OpenMP and C++ Threads into another plot shown in Figures \ref{fig:GPU} and \ref{fig:CPU}. In all figures we have plotted mean FLOPS with $95\%$ confidence intervals from the $2.5$ and $97.5$ percentiles as sampled by the bootstrap method. See Appendix \ref{appendix:Performance} for a detailed table of the metrics from these performances.

These plots with confidence intervals give strong evidence that some matrix multiplication algorithms have better performance. However, a formal statistical test give our evidence validity as it relates to our hypothesis. In Subsection \ref{subsection:Experiment} we stated that our hypothesis is that some algorithms have better performance than others. Formally, our null hypothesis is that the sample mean of FLOPS is the same for all algorithms. Our alternative hypothesis is that the sample mean is different for some algorithms. We take our largest matrix size $N = 10,000$ and perform formal statistical tests to determine whether some algorithms perform better than others.

To test for statistical significance we can analyze whether the sample means of performance as measured in FLOPS are different by using the significance level $0.01$. Since we have used the bootstrap method for resampling to analyze sample means of performance then the probability distributions of these averages is approximately normal by the central limit theorem (CLT). If the sample standard deviations for all distributions are equal then we can use One-Way ANOVA and Tukey tests to compare the sample means of our performances. However, Table \ref{tab:StdDevs} shows that the sample standard deviations of our distributions are not equal. Therefore, we can use Welch ANOVA and Games-Howell tests that do not make assumptions about the homogeneity of standard deviations across our distributions.

\begin{table}
\centering
\begin{tabular}{|c|c|c|c|c|}
\hline
CuBLAS & CUDA & BLAS & OpenMP & C++ Threads \\
\hline
5273.557 & 880.095 & 1213.323 & 21.177 & 1.306 \\
\hline
\end{tabular}
\caption{Sample standard deviations from matrix multiplication performance as measured in FLOPS where matrix size equals $10,000$.}
\label{tab:StdDevs}
\end{table}

A Welch ANOVA test for the sample mean FLOPS across all our matrix multiplication algorithms tells us that the p-value is $0.0$ that all our algorithms have the same performance. Therefore, we formally reject our null hypothesis and perform a Games-Howell test for our alternative hypothesis. Our subsequent Games-Howell test tells us with statistical significance that all algorithms have different performances as measured in sample mean FLOPS. Table \ref{tab:Pvalues} shows the p-values for all pairwise algorithms. Given our largest matrix size $N = 10,000$ and the sample mean FLOPS, we can formally order our matrix multiplication algorithms by performance from best to worst as CuBLAS, CUDA, BLAS, OpenMP, and C++ Threads.

\begin{table}
\centering
\begin{tabular}{|c|c|c|}
\hline
Group A & Group B & P-value \\
\hline
BLAS & C++ Threads & 0.000e+00 \\
BLAS & CuBLAS & 4.250e-12 \\
BLAS & CUDA & 0.000e+00 \\
BLAS & OpenMP & 2.751e-12 \\
C++ Threads & CuBLAS & 3.674e-12 \\
C++ Threads & CUDA & 0.000e+00 \\
C++ Threads & OpenMP & 0.000e+00 \\
CuBLAS & CUDA & 0.000e+00 \\
CuBLAS & OpenMP & 0.000e+00 \\
CUDA & OpenMP & 2.398e-13 \\
\hline
\end{tabular}
\caption{Statistically significant Games-Howell test p-values.}
\label{tab:Pvalues}
\end{table}

\section{Discussion}

There are fundamental differences in the designs of CPUs and GPUs that prevent CPUs from being as fast as GPUs in arithmetic heavy workloads like matrix multiplication \cite{hwu-2022}. CPUs are general purpose processors that must satisfy requirements from operating systems, applications, and I/O devices. In CPUs, arithmetic logic units (ALUs) are designed for latency, large caches are used to capture frequently accessed data, and sophisticated branch prediction and control logic are used at the cost of increased use of chip area. GPUs on the other hand are designed to perform a massive number of floating point operations per second for applications like deep learning, scientific simulations, and video graphics.

GPUs are designed to be single instruction multiple data (SIMD) processors \cite{hwu-2022}. Streaming processors (SPs) inside SMs are grouped into processing blocks where every eight cores form a processing block and share an instruction fetch/dispatch unit. This design allows for a smaller percentage of the hardware to be dedicated to control and larger percentage to be dedicated to arithmetic throughput. Although multiple processors are controlled by the same instructions each processor has its own registers for data.

\begin{figure}[h!]
    \centering
    \includegraphics[width=0.7\textwidth]{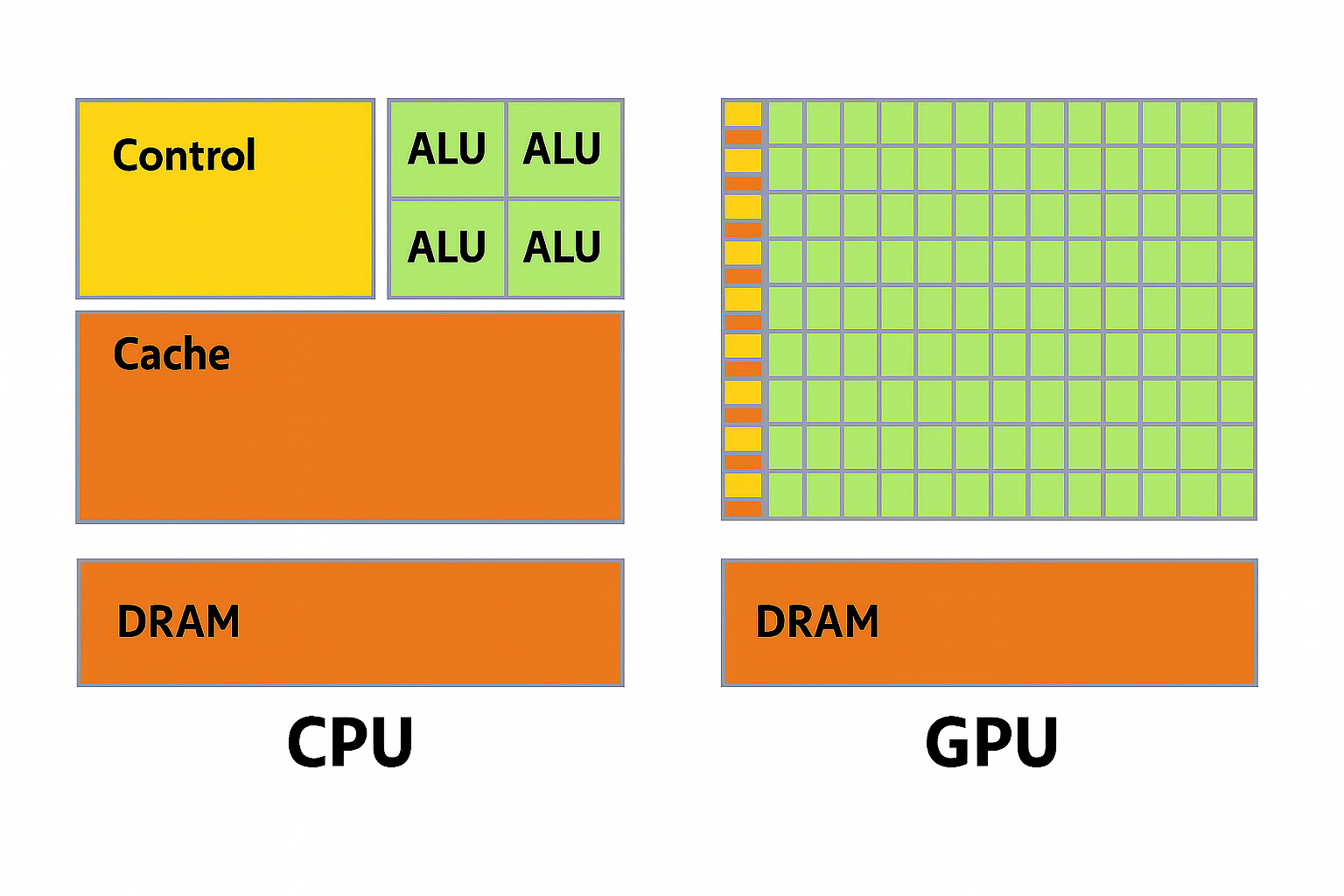}
    \caption{Latency versus throughput oriented design.}
    \label{fig:Architecture}
\end{figure}

These fundamental differences in chip design explain why our GPU bases matrix multiplication algorithms outperformed our CPU based algorithms. CuBLAS and BLAS are specialized for matrix operations and that explains their relative order of performance among our matrix multiplication algorithms. OpenMP is highly optimized for thread management which explains why it outperformed our C++ Threads algorithm.

All algorithms except CuBLAS reached their performance limit in FLOPS by our largest matrix size $N = 10,000$. This makes our experiment valid because our goal was to measure the performance of matrix multiplication algorithms. For larger sizes of $N$, this order of algorithms by performance in FLOPS would be fixed. Our hypothesis, that some algorithms perform better than others, was correct. With statistical significance we ordered our matrix multiplication algorithms by their performance.

\appendix

\section{Performance} \label{appendix:Performance}

\begin{figure}[h!]
    \centering
    \includegraphics[width=0.9\textwidth]{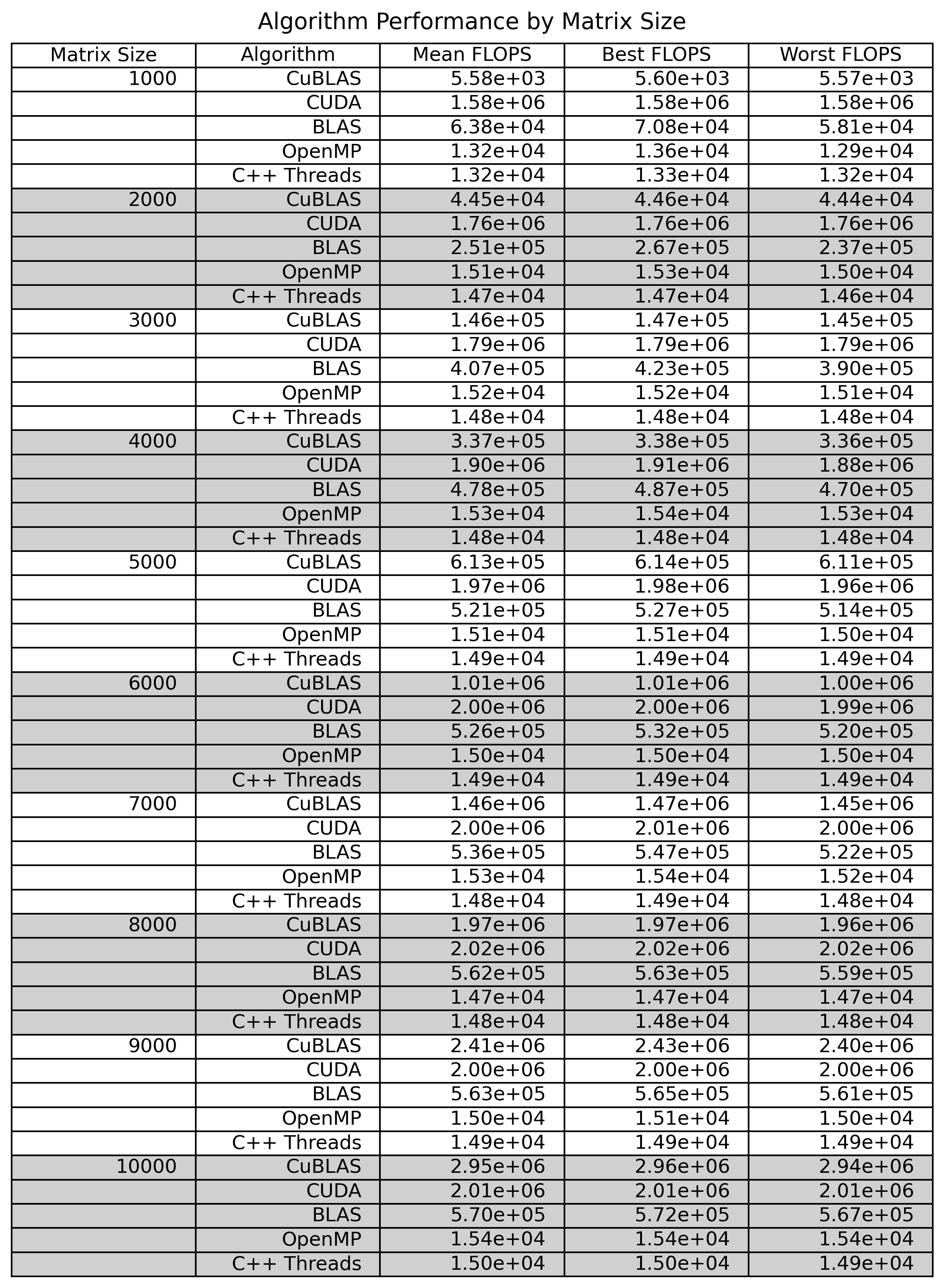}
    \caption{Performance with confidence intervals grouped by matrix size.}
    \label{fig:Performance}
\end{figure}

\printbibliography

\end{document}